\input{aipcheck}

\documentclass[
    ,final            
  ]
  {aipproc}

\layoutstyle{6x9}

\usepackage{amssymb}

\newcommand*{\Ave}[1]{\mathinner{\left\langle{#1}\right\rangle}}

\begin{document}

\title{Constraints on the two-Higgs-doublet model}


\author{Oscar St{\aa}l}{
  address={Department of Physics and Astronomy, Uppsala University\\ Box 516, SE-751 20 Uppsala, Sweden}
}

\begin{abstract}
 The two-Higgs-doublet model provides a simple, yet interesting, generalization of the SM Higgs sector. We study the CP-conserving version of this model with general, flavor-diagonal, Yukawa couplings. Indirect constraints are obtained from flavor physics on the charged Higgs boson mass and couplings. The relation of these bounds to those for the more specialized two-Higgs-doublet model types with a $Z_2$ symmetry is discussed.
\end{abstract}

\maketitle

\section{Introduction}
The two-Higgs-doublet model (2HDM) is the most popular extension of the Standard Model Higgs sector. It is theoretically well-motivated as the minimal Higgs sector compatible with supersymmetry (SUSY). In its general form, the 2HDM therefore serves as an effective theory for decoupled SUSY or other high-scale theories with two Higgs doublets present at low energy. Despite its simplicity the general 2HDM renders a wide range of new phenomena possible, such as CP-violation in the Higgs sector, FCNCs at tree-level, and the existence of a scalar charged current.

Lacking experimental evidence for either SUSY or the general 2HDM, the best option at hand is to use the available data to constrain the model parameters. Ideally this approach can also be used later, following a discovery, to discriminate between different models. Contributions of the charged Higgs boson $H^\pm$ to meson decays provides the most plausible low-energy signature of the 2HDM when tree-level FCNCs are absent. We therefore consider constraints from flavor physics on the couplings of $H^\pm$ to fermions, extending results for the 2HDM type II \cite{WahabElKaffas:2007xd,Flacher:2008zq} to more general coupling patterns. The reader is referred to \cite{Mahmoudi:2009zx} for the full analysis.

\section{Model and Constraints}
 We work here in a general, CP-conserving, 2HDM with two identical $\mathrm{SU(2)_L}$ doublets. The total number of free parameters in this model is large enough to accommodate the physical Higgs masses, including the charged Higgs boson mass $m_{H^\pm}$. A basis is chosen in the Higgs field space by specifying $\tan\beta\equiv\Ave{\Phi^0_2}/\Ave{\Phi^0_1}$.  The model is implemented in the general 2HDM code {\tt 2HDMC} \cite{Eriksson:2009ws}.
In all generality, the couplings of the charged Higgs boson to fermions are given by
\begin{equation}
\mathcal{L}=\overline{U}\left[\rho^{U}V_{\mathrm{CKM}}P_L-V_{\mathrm{CKM}}\rho^DP_R\right]DH^+-\overline{\nu}\rho^LP_R LH^++\mathrm{h.c.},
\end{equation}
where $U$, $D$, $L$ are vectors in flavor space and $\rho^F$ the corresponding $3\times 3$ Yukawa matrices. When all $\rho^F$ are diagonal the flavor structure of the charged Higgs couplings is fully described by the CKM matrix, while in the neutral sector tree-level FCNCs are absent. Introducing an ad-hoc $Z_2$ symmetry to achieve this leads to the well-known 2HDM ``types'' with $\rho_F=\sqrt{2}\lambda^FM^F/v$ proportional to the (diagonal) mass matrices $M^F$. In the type I model, $\lambda^U=\lambda^D=\lambda^L=\cot\beta$, while in type II $1/\lambda^U=-\lambda^D=-\lambda^L=\tan\beta$ instead. Two additional options exist where  leptons and down-type quarks couple differently. As an extension of the restrictive ``type'' models, we start from the ansatz \cite{Cheng:1987rs}
\begin{equation}
\rho^F_{ij}=\lambda^F_{ij}\frac{\sqrt{2m_im_j}}{v}\quad\mathrm{(no\ summation)},
\end{equation}
where $\lambda^F_{ij}$ are the free couplings. The smallness of the masses for the first generations leads to a natural suppression of off-diagonal elements for $\lambda$ of $\mathcal{O}(1)$. Given the phenomenological success of models without tree-level FCNCs, we make the stronger assumption that these are generically small enough to be neglected. Our focus is therefore to derive constraints on the free flavor-diagonal couplings $\lambda_{ff}$ (one for each fermion).

To constrain the $\lambda_{ff}$ and $m_{H^\pm}$, we determine the charged Higgs contribution to several flavor physics observables and compare to experimental data. The complete list of observables used is given in Table~\ref{tab:obs}.
For the numerical evaluation we use {\tt SuperIso} \cite{Mahmoudi:2007vz} (note that the SM values may deviate slightly from other values in the literature as a result of parametric updates). The main uncertainties in the theory predictions come from a few parameters: the light quark masses (which we take from the PDG), the top quark pole mass $\hat{m}_t=173.1\pm 0.6\pm 1.1$~GeV, the $B$ decay constant $f_B=190\pm13$~MeV \cite{Gamiz:2009ku}, and $|V_{ub}|=3.87\pm 0.09\pm 0.46$ \cite{Charles:2004jd}.
\begin{table}[b]
\begin{tabular*}{0.75\textwidth}{@{\extracolsep{\fill}}lrlr}
\hline
Observable & Experimental value & & Standard Model\\
\hline
$\rm{BR}(B\to X_s\gamma$) & $(3.52\pm 0.23\pm 0.09)\times 10^{-4}$ & \cite{Barberio:2008fa} & $(3.07\pm 0.22)\times 10^{-4}$ \\
$\Delta_0(B\to K^*\gamma$) & $(3.1\pm 2.3)\times 10^{-2}$ &\cite{Mahmoudi:2007vz} & $(7.8\pm 1.7)\times 10^{-2}$\\
$\Delta M_{B_d}$ (ps$^{-1}$) & $0.507\pm 0.004$ & \cite{Barberio:2008fa} & $0.53\pm 0.08$ \\
$\rm{BR}(B_u\to \tau\nu_\tau$) & $(1.73\pm 0.35)\times 10^{-4}$ & \cite{Charles:2004jd} & $(0.95\pm 0.27)\times 10^{-4}$ \\
$\xi_{D\ell\nu}(B\to D\tau\nu_\tau)$ &  $0.416\pm 0.117\pm 0.052$ & \cite{Aubert:2007dsa} & $0.297\pm 0.02$ \\
$R_{\ell 23}(K\to \mu\nu_\mu$) & $1.004\pm 0.007$ & \cite{Antonelli:2008jg} & $1$  \\
$\rm{BR}(D_s\to \mu\nu_\mu$) & $(5.8\pm 0.4)\times 10^{-3}$ & \cite{Akeroyd:2009tn} & $(4.98\pm 0.15)\times 10^{-3}$\\
$\rm{BR}(D_s\to \tau\nu_\tau$) & $(5.7\pm 0.4)\times 10^{-2}$ & \cite{Akeroyd:2009tn} & $(4.82\pm 0.14)\times 10^{-2}$ \\
\hline
\end{tabular*}
\caption{Observables used to constrain the charged Higgs boson in the 2HDM. SM values are evaluated using {\tt SuperIso} \cite{Mahmoudi:2007vz}.}
\label{tab:obs}
\end{table}

\section{Results and Conclusions}
\begin{figure}
\begin{tabular}{cc}
\includegraphics[width=0.42\columnwidth]{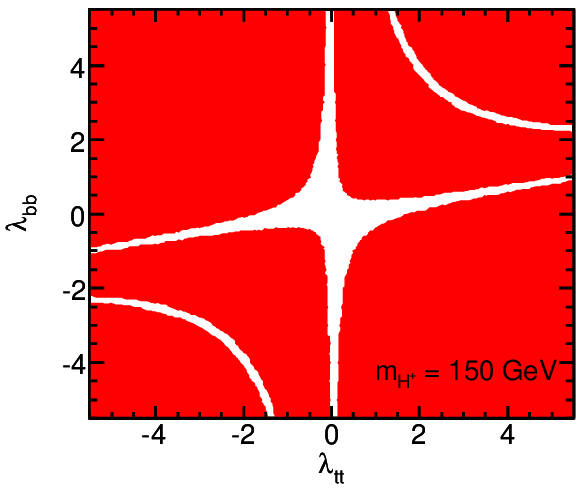} &
\includegraphics[width=0.42\columnwidth]{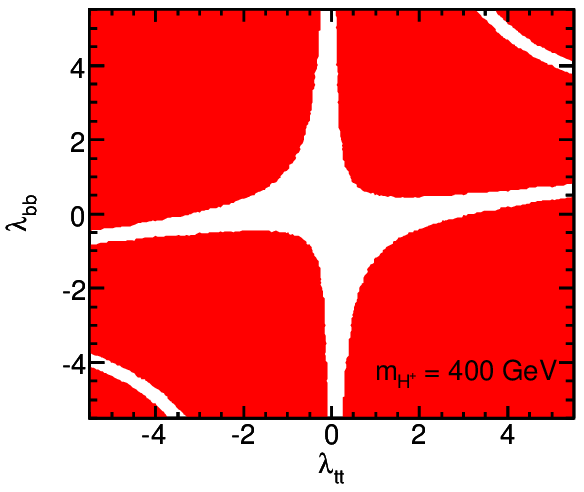} \\
\end{tabular}
\caption{Excluded regions (in color) at $95\%$ CL from $\mathrm{BR}(B\to X_s\gamma)$ for $m_{H^\pm}=150$~GeV (left) and $m_{H^\pm}=400$~GeV (right).}
\label{fig:bsgamma}
\end{figure}

\begin{figure}
\begin{tabular}{cc}
\includegraphics[width=0.42\columnwidth]{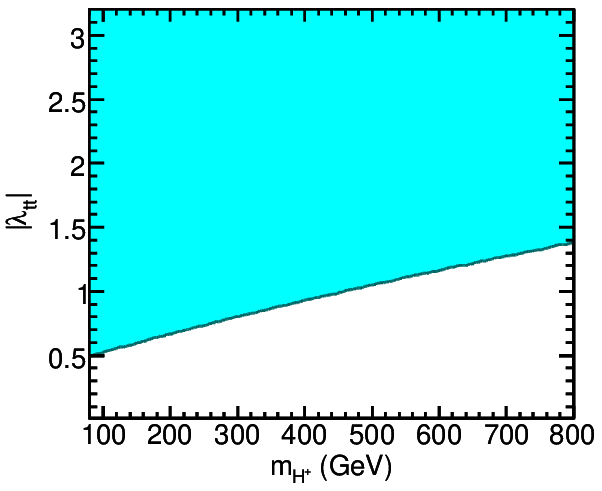} &
\includegraphics[width=0.42\columnwidth]{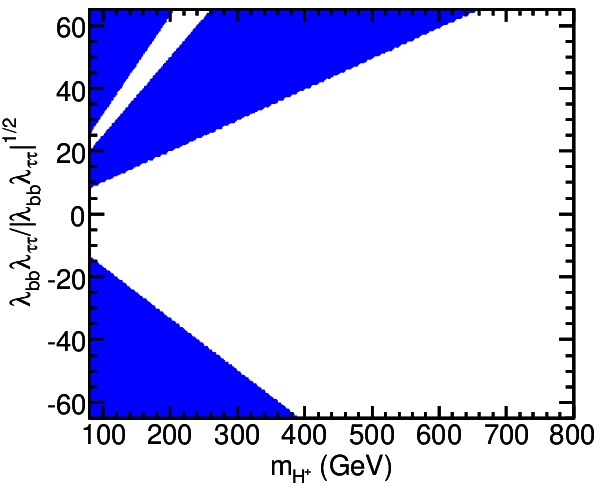} \\
\end{tabular}
\caption{Excluded regions (in color) at $95\%$ CL from $\Delta M_{B_d}$ (left) and $\mathrm{BR}(B_u\to\tau\nu_\tau)$ (right).}
\label{fig:dmb}
\end{figure}

For each observable separately we determine the excluded parameter regions at $95\%$ CL  assuming Gaussian errors, one degree of freedom. Masses $m_{H^\pm}< 80$~GeV, which are excluded by LEP \cite{Abdallah:2003wd}, are not considered. In Figure~\ref{fig:bsgamma} the results for the loop-mediated decay $B\to X_s\gamma$ are shown.
Both for $m_{H^\pm}=150$ and $400$ GeV only small regions of the $\left(\lambda_{tt}, \lambda_{bb}\right)$ parameter space are allowed. However, interesting regions still exist with small $|\lambda_{tt}|\ll 1$ and large $|\lambda_{bb}|$. Since $m_t$ is large, a small value $\lambda_{tt}\lesssim 1$ is favored by arguments of perturbativity.
In Figure~\ref{fig:dmb}, the left plot shows the region excluded from the mass difference $\Delta M_{B_{d}}$ measured in $B\overline{B}$ oscillations. The leading 2HDM contribution -- from a box diagram with $H^\pm$ and the top quark -- is proportional to $|\lambda_{tt}|^4$. As a result of this strong dependence, we find that $|\lambda_{tt}|>1$ is excluded for $m_{H^\pm}\lesssim 500$~GeV.
The right plot in Figure~\ref{fig:dmb} shows constraints from $B_u\to\tau\nu_\tau$ which is mediated by the charged Higgs at tree-level. From the figure we see that large values of the product $|\lambda_{bb}\lambda_{\tau\tau}|$ are excluded. For large and positive $\lambda_{bb}\lambda_{\tau\tau}$, a cancellation occurs between the $H^\pm$ and SM contributions, resulting in an allowed region.
In addition to the constraints presented here, ref.~\cite{Mahmoudi:2009zx} contains results for each of the observables listed in Table~\ref{tab:obs}.

\begin{figure}
\begin{tabular}{cc}
\includegraphics[width=0.41\columnwidth]{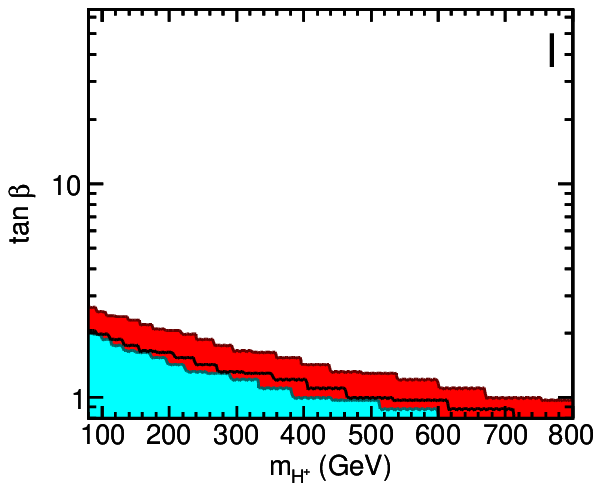} &
\includegraphics[width=0.41\columnwidth]{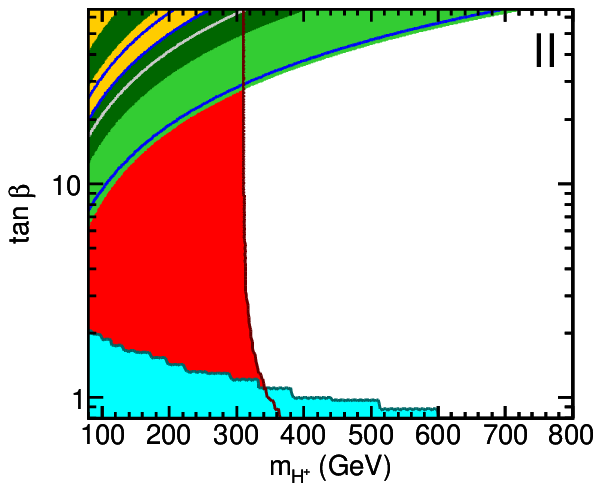} \\
\end{tabular}
\caption{Constraints on 2HDM types I and II. Colors corresponds to constraints at $95\%$ CL from $\mathrm{BR}(B\to X_s\gamma)$ (red), $\Delta_0(B\to K^*\gamma)$ (black contour), $\Delta M_{B_d}$ (cyan), $\mathrm{BR}(B_u\to\tau\nu_\tau)$ (blue), $\mathrm{BR}(B\to D\tau\nu_\tau)$ (yellow), $R_{\ell 23}(K\to\mu\nu)$ (grey contour), $\mathrm{BR}(D_s\to\tau\nu_\tau)$ (light green), and $\mathrm{BR}(D_s\to\mu\nu_\mu)$ (dark green).}
\label{fig:Z2}
\end{figure}

Since the number of useful observables is limited, a combination of the constraints usually requires making universality assumptions on the Yukawa couplings. To this end, Figure~\ref{fig:Z2} shows combined constraints on the 2HDM types I and II. In the figure exclusion regions from different observables (each corresponding individually to $95\%$ CL) are superimposed.
We find that low values of $\tan\beta\lesssim 1$ are excluded for $m_{H^\pm}<500$~GeV in both types as a result of $\Delta M_{B_d}$ and $b\to s\gamma$ transitions. For the 2HDM type II, we observe the lower limit $m_{H^\pm}\gtrsim 300$~GeV arising from $\mathrm{BR}(B\to X_s\gamma)$. This $\tan\beta$-independent limit results from a contribution proportional to $\lambda_{tt}\lambda_{bb}\sim 1$.
Most of the observables investigated only constrain the type II model at high $\tan\beta$. The relevant contribution of $H^\pm$ to the leptonic and semi-leptonic decay modes is proportional to the product $\lambda_{bb}\lambda_{\ell\ell}$ (or similarly for the second generation), which is only sufficiently enhanced (as $\tan^2\beta$) in the type II model. Processes involving leptons exclude charged Higgs masses $m_{H^\pm}\lesssim 500$~GeV for $\tan\beta\gtrsim 60$ in the case of type II couplings.

In conclusion, the charged Higgs contribution to flavor physics observables is interesting to distinguish e.g. the MSSM from a more general 2HDM. We have therefore presented constraints on the couplings from each observable separately without any universality assumptions. For models with a $Z_2$ symmetry, the popular 2HDM type II is severely constrained, with $b\to s\gamma$ leading to the general limit $m_{H^\pm}\gtrsim 300$~GeV; an even stronger limit is obtained both for high and low $\tan\beta$. For the type I model, no limit is obtained on $m_{H^\pm}$ beyond that from LEP. On the other hand, $\tan\beta\gtrsim 1$ is required for $m_{H^\pm}\lesssim 800$~GeV,  leading effectively to a decoupling of the quark sector.

I thank F.~Nazila Mahmoudi for the collaboration on this project.

\bibliographystyle{aipproc}   
\bibliography{os_susy09}

\end{document}